%% file: main.tex
\documentclass[prl,amsmath,amssymb,aps,showpacs,10pt,superscriptaddress,twocolumn]{revtex4-1}
\usepackage[utf8]{inputenc}
\usepackage{amsmath}
\usepackage{latexsym}
\usepackage{amsfonts}
\usepackage{epsfig}
\usepackage{psfrag}
\usepackage{graphicx}
\usepackage{hyperref}
\usepackage{color}
\usepackage{soul,xcolor}
\usepackage[normalem]{ulem}

\definecolor{black}{rgb}{0,0,0}
\definecolor{blue}{rgb}{0,0,1}
\definecolor{green}{rgb}{0,1,0}
\definecolor{red}{rgb}{1,0,0}
\definecolor{brown}{rgb}{0.4,0.2,0}
\definecolor{darkgreen}{rgb}{0,0.7,0}

\renewcommand{\vec}[1]{\boldsymbol #1}
\newcommand{\ket}[1]{\left|#1\right>}
\newcommand{\bra}[1]{\left<#1\right|}

\newcommand{\bea}{\begin{eqnarray}}
\newcommand{\ea}{\end{eqnarray}}
\newcommand{\eea}{\end{eqnarray}}

\newcommand{\sumint}[1]
{\begin{array}{c} \\
{{\textstyle\sum}\hspace{-1.1em}{\displaystyle\int}}\\
{\scriptstyle{#1}}
\end{array}}

\usepackage[percent]{overpic}
 \newlength{\imagewidth}

\begin{document}
\setlength{\imagewidth}{1\linewidth}

\title{Effective three-body interactions in Cs($6s$)-Cs($nd$) Rydberg trimers
}
\author{Christian Fey}
\affiliation{Zentrum f\"ur optische Quantentechnologien, Universit\"at Hamburg, Luruper Chaussee 149, 22761 Hamburg, Germany}
\affiliation{ITAMP, Harvard-Smithsonian Center for Astrophysics 60 Garden St., Cambridge, Massachusetts 02138, USA}
\author{Jin Yang}
\affiliation{Homer L. Dodge Department of Physics and Astronomy, The University of Oklahoma, Norman, Oklahoma 73072, USA}
\author{Seth T. Rittenhouse}
\affiliation{Department of Physics, The Naval Academy, Annapolis, Maryland 21402, USA}
\author{Fabian Munkes}
\author{Margarita Baluktsian}
\affiliation{Homer L. Dodge Department of Physics and Astronomy, The University of Oklahoma, Norman, Oklahoma 73072, USA}
\author{Peter Schmelcher}
\affiliation{Zentrum f\"ur optische Quantentechnologien, Universit\"at Hamburg, Luruper Chaussee 149, 22761 Hamburg, Germany}
\affiliation{The Hamburg Centre for Ultrafast Imaging, Universit\"at Hamburg, Luruper Chaussee 149, 22761 Hamburg, Germany}
\author{H. R. Sadeghpour}
\affiliation{ITAMP, Harvard-Smithsonian Center for Astrophysics 60 Garden St., Cambridge, Massachusetts 02138, USA}
\author{James P. Shaffer}
\affiliation{Homer L. Dodge Department of Physics and Astronomy, The University of Oklahoma, Norman, Oklahoma 73072, USA}

\date{\today}

\begin{abstract}
Ultralong-range Rydberg trimer molecules are spectroscopically observed in an ultracold gas of Cs($nd_{3/2}$) atoms. The atomic Rydberg state anisotropy allows for the formation of angular trimer states, whose energies {\it may not} be obtained from integer multiples of dimer energies. These nonadditive trimers are predicted to coexist with Rydberg dimer lines. The existence of such effective three-body interactions is confirmed  with observation of asymmetric line profiles and interpreted by a theoretical approach which includes relativistic spin interactions. Simulations of the observed spectra with and without angular trimer lines lend convincing support to the existence of effective three-body interactions.
\end{abstract}

\maketitle

Ultralong-range Rydberg molecules (ULRM) form when a Rydberg electron scatters from a ground state atom within the Rydberg orbit\cite{greene_creation_2000,bendkowsky_observation_2009}. These molecules are interesting for their striking features, such as an exotic binding mechanism, huge sizes ($\sim$ 100 nm), and large permanent electric dipole moments ($\sim$ kilo-Debyes) \cite{booth_production_2015}, as well as for quantum many-body phenomena at high densities \cite{gaj_molecular_2014,balewski_coupling_2013,schlagmuller_probing_2016,schlagmuller_ultracold_2016,eiles_ultracold_2016,camargo_creation_2018, luukko_polyatomic_2017}. Following the original prediction of ultralong-range Rydberg diatomic molecules \cite{greene_creation_2000}, theoretical work predicted the formation of polyatomic ULRM where on average more than one perturber is bound within the Rydberg electron orbit \cite{liu_polyatomic_2006,liu_ultra-long-range_2009}.
It is appropriate to distinguish these molecules by the angular momentum $l$ of the Rydberg electron.
Triatomic molecules with $l=0$ were observed shortly after the initial discovery of diatomic ULRM \cite{bendkowsky_rydberg_2010}. 
Additionally, observations of similar and higher order Rb and Sr Rydberg oligomers have been reported \cite{gaj_molecular_2014,schlagmuller_probing_2016,camargo_lifetimes_2016,camargo_creation_2018}. Since the Rydberg electron is in an isotropic $s$-state, the interactions among the nuclei are additive and the binding energies are always integer multiples of the dimer energies.

\begin{figure*}[ht]
\includegraphics[width= 0.9 \textwidth]{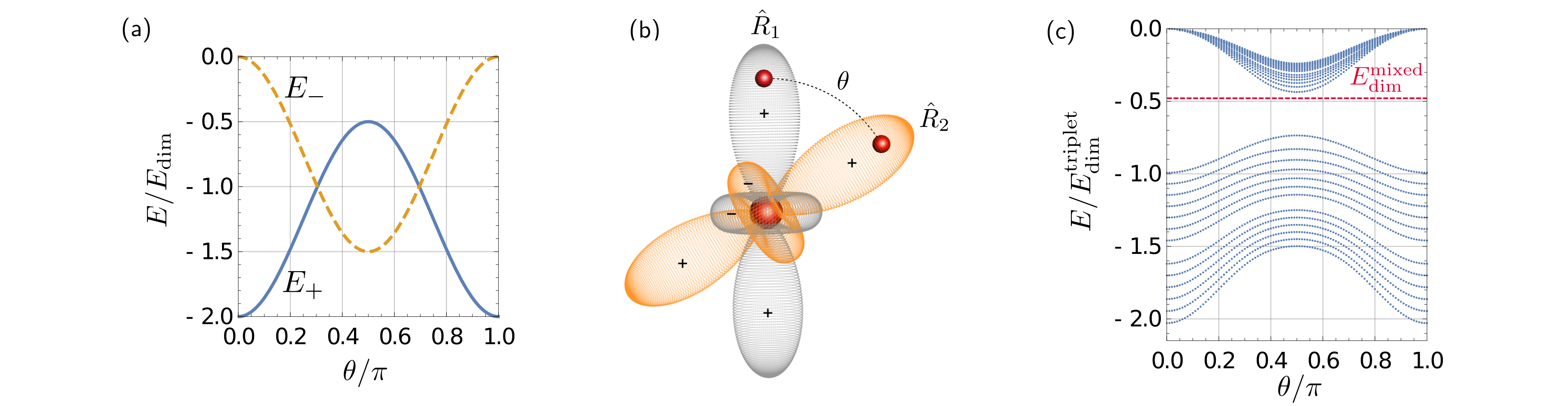}
\caption{The angular trimer potential energy curves (a) for $R_1=R_2=1868$ a$_0$ for excitation into Cs($34d_{3/2}$) Rydberg state, without any spin-dependent terms in the interaction Hamiltonian, i. e. $E_{\pm}$ terms in equation (\ref{eqn:E_nospinp}) rescaled by $E_\text{dim}(R=1868 \, a_0)$. The primitive orbitals (signs indicated) whose superpositions $\psi_\pm(\vec{r};\vec{R}_1;\vec{R}_2)$ produce the eigenvalues $E_{\pm}$, are illustrated in (b). The angular configuration of the Rydberg core (at the origin) and the two ground state atoms (unit vectors $\hat{R}_1$, $\hat{R}_2$) are represented by the central sphere and the outer spheres, respectively. In (c) the relativistic spin-dependent interactions deform the curves $E_\pm$ and lift degeneracies, leading to modified angular potentials. Here $E^\text{triplet}_\text{dim}=71$ MHz ($E^\text{mixed}_\text{dim}=34$ MHz ) corresponds to the energy at the minimum of the outer-most well of the triplet dominated (mixed singlet/triplet) dimer PES, see dashed blue (red) curve in Fig.  \ref{fig2}.}
\label{fig1}
\end{figure*}

In this work, we report on the observation of a new class of ultralong-range Rydberg molecules which form primarily due to the {\it intrinsic anisotropy} of Cs($nd)$ Rydberg wave functions, see Fig.~\ref{fig1}. These Rydberg trimer states appear in the experiment red  detuned from the atomic lines, when Cs atoms are photo-excited into $nd_{3/2}$ Rydberg states. 
Computations of Born-Oppenheimer (BO) potential energy surfaces (PES) enable us to simulate theoretical line profiles for dimer and trimer signals which are compared to the experimental spectra and lend support to the existence of non-additive effective three-body interactions in the molecular Hamiltonian.
A crucial aspect of the observed three-body spectral profile is the influence of relativistic  spin-orbit and hyperfine interactions which lift the underlying degeneracy in the non-relativistic molecular Hamiltonian. Strong support for our interpretation is found in the simulated spectra which include only dimer states vs the spectra which combine dimer and trimer states. These non-additive trimers coexist with Rydberg dimer lines. 

The formation of Rydberg trimers whose energies are {\it not} combinatorially derived from dimer energies is a bound molecular realization of an effective three-body interaction whose existence is not manifest in pair-wise interactions. Effective three-body potentials, as products of two-body interactions, have been phenomenologically derived from the vibrational spectra of alkali metal trimers \cite{Higgins1996}. Three-body and higher multi-body interactions appear in various models for many-body condensed matter and quantum information Hamiltonians, including the quantum loop models describing topological order with four spins \cite{Kitaev2003}, adiabatic quantum computing Hamiltonians \cite{Aharonov2007}, and perturbation gadget Hamiltonians \cite{Kempe2006}. 
Effective three-body Hamiltonians have been proposed in atomic and molecular systems with polar molecules in optical lattices \cite{Buechler2007}, for three-interacting bosons near a two-body Feshbach resonance \cite{Petrov2014}, with optical lattice modulation \cite{Daley2014}, and circuit QED systems \cite{Hafezi2014}. Three-body F\"orster resonances, where two-body resonances are absent, were recently observed in Rydberg excitation in an ultracold Rb gas \cite{Faoro2015, tretyakov_observation_2017}.

In its simplest manifestation, the interaction of a Rydberg electron at position $\vec{r}$ and spin $\vec{s}_{r}$  with two ($i=1,2$) ground state alkali metal atoms at positions $\vec{R}_i$ and spins $\vec{s}_{i}$ can be described by a Fermi pseudopotential \cite{fermi_sopra_1934}. In atomic units and for pure $s$-wave scattering, this interaction potential is given by $\hat{V}=\hat{V}_1+ \hat{V}_2$, where

\begin{align}
\begin{split}
\hat{V}_{i}& =2 \pi  \delta (\vec{r}-\vec{R}_{i}) \left[ a^S_s(k_{i}) \hat{P}^S_{i} + a^T_s(k_{i}) \hat{P}^T_{i}\right].
\label{eqn:pseudopotential}
\end{split}
\end{align}
The operators $\hat{P}^{S(T)}_{i}$ project independently onto the singlet (S) and triplet (T) spin channels of the two alkali metal atoms, where $\hat{P}^T_{i}= \vec{s}_{r}\cdot \vec{s}_{i} +3/4 $ and $\hat{P}^S_{i}=1 -\hat{P}^T_{i} $. The interaction strength in each channel is determined by the singlet (triplet) $s$-wave scattering length $a^{S(T)}_s(k_i)$ for a ground state atom and a free electron with momentum $k_{i}$.

\begin{figure}[ht]
\includegraphics[width= 0.99 \linewidth]{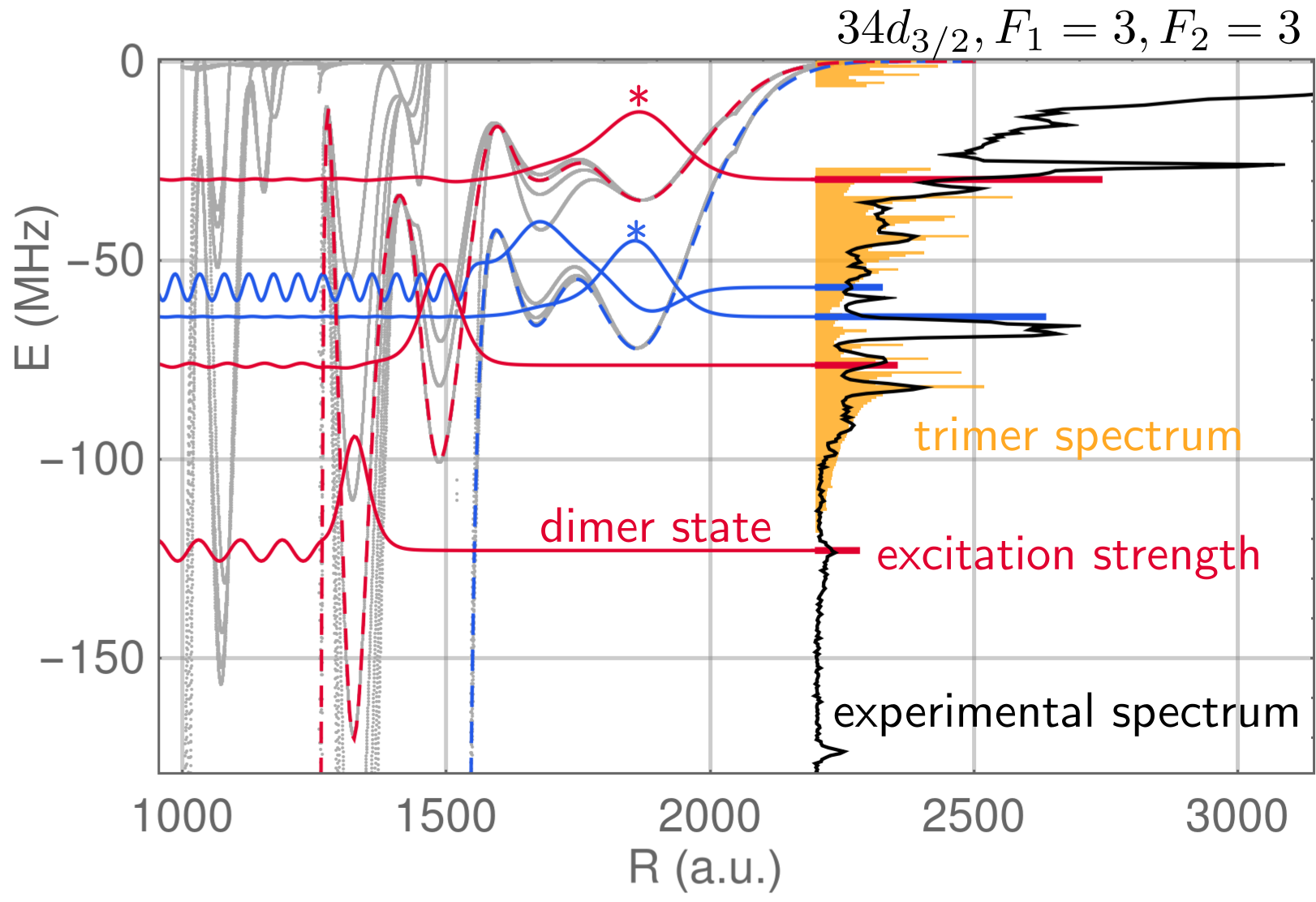}
\caption{Comparison between the experimental spectrum (black line) and theoretical results for molecules close to the Cs($34d_{3/2}$) $F_1=F_2=3$ dissociation energy. Vibrational wave functions (colored full lines) are presented for a few diatomic potential curves (dashed lines) that have been selected from all potential energy curves (gray background). The offset of the wave function corresponds to its vibrational energy. The length of the colored bars on top of the experimental spectrum are the theoretical line strengths. The computed trimer histogram (bin width 700 KHz) is superimposed in orange. Dimer states localized in the most outer wells are marked with a star.}
\label{fig2}
\end{figure}

When a ground state atom lies within the Rydberg electron cloud, the contact potential perturbs the Rydberg energy levels and leads to oscillatory PES \cite{greene_creation_2000, bendkowsky_observation_2009}. These PES are depicted in Fig. \ref{fig2} for a Cs($34d_{3/2}$) Rydberg electron where only the first ground state atom (hyperfine state $F_1=3$) is present, i.e. $R_2 \to \infty$ and $R:=R_1$. For large distances, $R>1800$ a$_0$ with a$_0$ the Bohr radius, one finds a set of deep potentials (dashed blue line) with almost pure electronic triplet character and a set of shallow potentials (dashed red line) with mixed singlet/triplet character \cite{anderson_angular-momentum_2014,anderson_photoassociation_2014}. This spin mixing results from the interplay of three competing interactions: the Fermi pseudopotential, the Rydberg electron spin-orbit coupling and the ground state atom hyperfine interactions. Vibrational dimer states bound in the outer wells (similar to the colored solid lines) have been confirmed experimentally \cite{anderson_photoassociation_2014, sasmannshausen_experimental_2015,bottcher_observation_2016}. At distances $R<1800$ a$_0$, additional $p$-wave interactions become important \cite{hamilton_shape-resonance-induced_2002, khuskivadze_adiabatic_2002, eiles_hamiltonian_2017}. They lift the degeneracy of the potentials and lead to sharp drops at distances where the Rydberg electron $p$-wave scattering phase shifts are resonant. These effects are described in more detail in \cite{supp}.

When two ground state atoms lie within the electronic Rydberg cloud, the PESs will depend on $R_1$, $R_2$ and the enclosed angle $\theta$, see Fig. \ref{fig1}.
In the absence of all spin interactions, analytical expressions for PES can be derived. This has been demonstrated for triatomic trilobites \cite{liu_polyatomic_2006,liu_ultra-long-range_2009} as well as for triatomic low-angular momentum states \cite{fey_stretching_2016}. For $R_1=R_2$, the $l=2$ trimer eigenstates can be expressed as linear combinations $\psi_\pm(\vec{r};\vec{R}_1;\vec{R}_2)= \mathcal{N}_\pm(\vec{R}_1,\vec{R}_2) \left[\psi_\text{dim}(\vec{r};\vec{R}_1)\pm \psi_\text{dim}(\vec{r};\vec{R}_2) \right]$ of the dimer eigenstates $\psi_\text{dim}(\vec{r};\vec{R})= \sum_m  \phi_{n,l=2,m}(\vec{R})^* \phi_{n,l=2,m}(\vec{r})$, with normalization  $\mathcal{N}_\pm(\vec{R}_1,\vec{R}_2)$ and Rydberg wave functions $\phi_{n,l=2,m}(\vec{r})$. The corresponding angular potentials are
\begin{equation}
E_\pm(R,\theta)= E_\text{dim}(R) \left[1 \pm \left(-\frac{1}{2}+ \frac{3}{2} \cos^2 \theta\right)\right] \label{eqn:E_nospinp},
\end{equation}
where $E_\text{dim}(R)$ is the corresponding diatomic PES.

This result is illustrated in Fig. \ref{fig1} (a). The equilibrium angles are $\theta=0$ and $\theta=\pi$ for $E_+$ and $\theta=\pi/2$ for $E_-$. These angles are energetically favored since the electronic wave function maximizes its density on the two ground state atoms in those configurations, see Fig. \ref{fig1} (b). The minima in the $E_+$ channel give twice the binding energy of the dimer state and this is the case which has thus far been discussed in formation of bound Rydberg trimers. The minimum of the $E_-$ channel in the perpendicular configuration ($\theta=\pi/2$) is particularly interesting because it is not a global minimum, but yet can support vibrational states at 3/2 the dimer binding energy. The above two-state model serves to highlight an essential finding of this work, namely the {\it prediction of angular Rydberg trimer states in the $E_-$ channel detuned to the red of the Cs($nd_{3/2})$ thresholds.}

We next turn on the relativistic spin mixing interactions for the Rydberg electron and the two ground state atoms. Relevant are the spin-orbit coupling $\vec{j}=\vec{s}_r+\vec{l}$, where $\vec{l}$ is the angular momentum of the Rydberg electron, as well as the hyperfine coupling $\vec{F}_i=\vec{I}_i + \vec{s}_i$ for each perturber ($i=1,2$). The nuclear spin $I$ of Cs is 7/2. As for the diatomic case, these interactions couple the different spin channels of the contact interaction in (\ref{eqn:pseudopotential}) and lift the $E_\pm$ degeneracy. This gives rise to the BO angular potential energy curves in  Fig.~\ref{fig1}(c).
Like in the simple two-state picture one finds a first set of shallower $E_-$-type curves with minima at $\theta=\pi/2$ and a second set of deeper $E_+$-type curves with minima at $\theta=0$ and $\theta=\pi$.

These PESs demonstrate that the Rydberg electron mediates an effective three-body-force between the three atomic cores.
The range of this interaction is on the order of the Rydberg orbit.
A detailed analysis of the transition from Fig. \ref{fig1} (a) to Fig. \ref{fig1} (c), where we trace back certain features of the angular structure to certain types of spin-interactions, is performed in \cite{supp}.

\begin{figure}[t]
\includegraphics[width= 0.49 \textwidth]{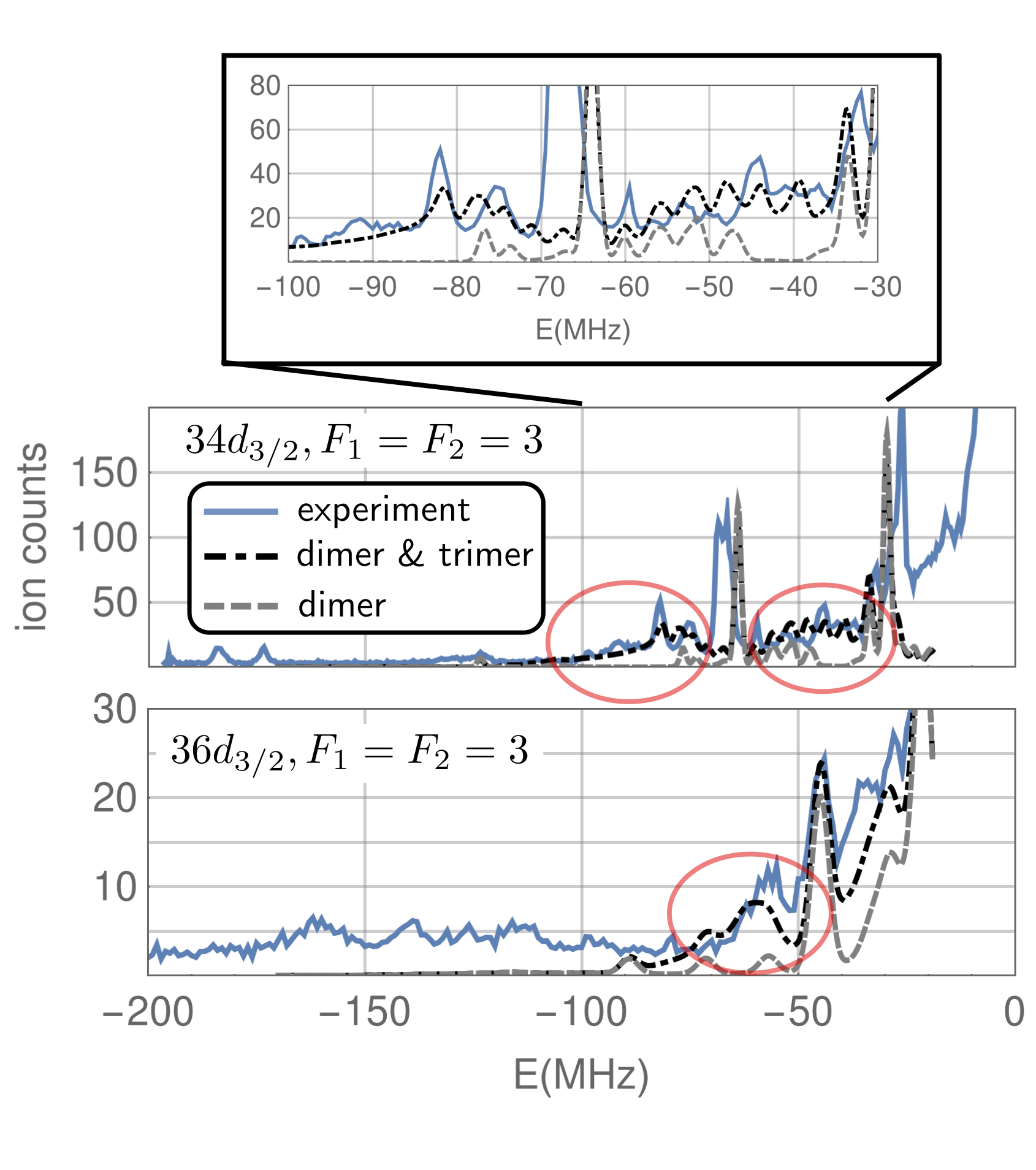}
\caption{Comparison of the experimental spectra (solid blue lines) and theoretical simulations including molecular dimer and trimer signals (dashed dotted black lines) close to the Cs($34d_{3/2}$ and $36d_{3/2}$) atomic Rydberg lines with $F_1=F_2=3$. To estimate the impact of trimer signals we present also theoretical simulations including only dimer states (dashed gray lines). The red circles highlight the energy regions where the contributions of the non-additive trimers are most prominent in signal height and energy spread. The asymmetric long shoulders of these peaks can be explained only by the non-additive trimer signals. The upper panel displays a magnification of the region with dominant trimer contributions for $34d_{3/2}$.} 
\label{fig3}
\end{figure}

The experiments were performed in a crossed far-off resonance trap (FORT). The maximum number density of the FORT is $\sim 1\times10^{13}$ cm$^{-3}$ at a temperature of $\sim 40~\rm{\mu}$K. The photoassociation of the $nd$ Rydberg molecules is achieved using a two-photon transition. Cs atoms are excited from the ground state, 6$s_{1/2}(F=3)$, to Rydberg states using a near resonant transition to 6$p_{3/2}(F=4)$ at $852\,$nm. This first step of the excitation is detuned from resonance by $300\,$MHz. The laser that drives the first step of the transition is locked to a Cs saturated absorption setup and the Rydberg excitation laser at $\sim 508\,$nm, is locked to a Fabry-P$\acute{\rm{e}}$rot cavity. Both lasers are linearly polarized in the same direction.
The overall stability of the laser systems for the Cs$(34d_{3/2})$ and Cs$(36d_{3/2})$ measurements is  $\sim 700$ KHz and $\sim 3\,$MHz, respectively. 
The higher resolution is achieved by locking the lasers to an ultra-stable reference cavity.
A signal is generated by ionizing the Rydberg atoms and projecting them onto a microchannel plate (MCP) detector in a time-of-flight spectrometer. The signal is collected as a function of laser detuning. Each prepared sample is excited with a
series of laser pulses for a time of $150\,$ms which amounts to $300\,$cycles of the excitation sequence. Each excitation pulse lasts $30~\mu$s and is followed by a $67\,$V$\,$cm$^{-1}$ electric field pulse which lasts for $500\,$ns. Rydberg molecules can produce both Cs$^{+}$ ions and Cs$_{2}^{+}$ ions.
For the Cs$(36d_{3/2})$ measurements we observe only the Cs$^{+}$ ions, whereas, for the Cs$(34d_{3/2})$ measurements, a better signal to noise ratio is achieved by observing both, Cs$^{+}$ and Cs$_{2}^{+}$.
The resulting spectra are presented in Fig.~\ref{fig2} and Fig.~\ref{fig3}.
Background electric fields in the experimental apparatus were measured to be $15\,$mV$\,$cm$^{-1}$.

To compare our electronic structure calculations with experimentally observed spectra, we follow two complementary approaches for dimer and trimer states.
For the Rydberg dimers, we compute vibrational wave functions $\chi_\nu(R)$, see Fig. \ref{fig2} for the $34d_{3/2}$ excitation, with a finite difference method.
The calculation yields bound states, including those states bound by quantum reflection from steep potential drops \cite{bendkowsky_rydberg_2010}. To estimate the line strengths $\Gamma_\nu$, we compute the Franck-Condon factors as
$\Gamma_\nu \propto \left| \int dR \chi_0(R) R^2 \chi_\nu(R) \right|^2 $, where the initial state $\chi_0(R)$ is the pair wave function for two ultracold atoms in the ground state, here a constant \cite{desalvo_ultra-long-range_2015}.

Calculated dimer states (colored full lines) are presented in Fig. \ref{fig2} together with their line strengths (length of the colored horizontal bars) and are compared with the observed spectrum (black line). For ease of readability, we present only the most prominent dimer states with relatively large line strengths from a subset of potential curves (colored dashed curves), selected from all BO potentials (remaining gray curves). Since the line strength scales approximately quadratically with the bond length, the states in the most outer wells dominate the spectrum. In Fig. \ref{fig2} these states are marked by stars and can be clearly identified with the strongest molecular lines in the experiment.

In a more detailed analysis, we use the computed dimer states as an input for a simulation of the dimer spectrum.
The set of dimer states comprises not only those shown in Fig.~\ref{fig2} but all of the dimer states from the BO potentials that connect to the $34d_{3/2}$ atomic asymptote, independently of the strength of their Franck-Condon factors.
In accordance with the experimental parameters, we generate these spectra by convoluting the discrete line strengths with Gaussian profiles having a width of 700 KHz (3 MHz) for $34d_{3/2}$ ($36d_{3/2}$).
These simulations are presented in Fig.~\ref{fig3} (gray dashed lines) and are compared to the experimental signals (solid blue lines). The agreement of the line strengths and profiles is reasonable. However, some prominent spectral features are not captured with the dimer calculations, e.g. the broad peaks near the 45 and 80 MHz detuning in the Cs($34d_{3/2}$) spectrum (red circles).  We show below that these features can {\it only} be explained with the addition of non-additive trimer states, demonstrating the existence and observation of an effective three-body interaction in the formation of trimer molecules.

Because the density of angular states in the Rydberg molecules is large, we use a sampling technique to obtain line profiles of trimer states. This differs from the method used in Ref. \cite{fey_stretching_2016}. 
Our approach is motivated by the fact that the trimer spectrum is dominated by signals from molecules where both ground state atoms are in the most outer wells, e.g. $R_1=R_2=1868 \text{a}_0$ for $34d_{3/2}$, and that the experimental laser linewidth is larger than the spacing of bending excitations. 
Experimental signals of trimers with shorter bond lengths are suppressed, since the probability to find two ground-state atoms at distances $R_1$ and $R_2$ from the Rydberg atom scales as $(R_1 R_2)^2$.
The trimer signal is then obtained as a histogram of energies $E(R_1,R_2,\theta)$, generated by drawing random molecular configurations with fixed $R_1=R_2$ and variable $\theta$. 
For $34d_{3/2}$ ($36d_{3/2}$) we use $R_1=R_2=1868 \text{a}_0$ ($R_1=R_2=2110 \text{a}_0$) and adjust the bin width to the 700 KHz (3 MHz) spectroscopic resolution.
Furthermore we assume that the relative distribution of the atoms is isotropic (which implies a $\sin\theta$ distribution for the sampling) and take into account all the energy surfaces shown in Fig.~\ref{fig1}~(c).
Hence, large contributions in the histogram are expected for energies close to $E(R_1,R_2,\theta \approx \pi/2)$, for stationary points in the PESs or for energetic regions where the PESs are very dense.
Since the molecule is frozen in a rigid rotor approximation to its radial equilibrium position, we account for the zero-point vibrational motion by blue-shifting the resulting histograms by 25 MHz (15 MHz) for $34d_{3/2}$ ($36d_{3/2}$). This sampling approach can be viewed as a modification of methods presented in \cite{schlagmuller_probing_2016} that were successfully applied to describe the spectrum of polyatomic ULRM in $s$-states.
The resulting histogram for $34d_{3/2}$ is displayed in orange in Fig. \ref{fig2} in comparison with the observed signal (for $32d_{3/2}$, $36d_{3/2}$ and $38d_{3/2}$, see \cite{supp}).
Many characteristics of the shape of the histograms (such as the relative heights and the asymmetry of the peaks) agree with the observations. 
Due to the anisotropy of the PES, the energy positions of the simulated trimers cannot be obtained by addition of dimer energies. For $34d_{3/2}$ the corresponding dimer states that localize at $R=1868 \text{a}_0$ are marked with a star in Fig. \ref{fig2}.

With the relatively high atomic density, we expect the experimental spectrum to contain both dimer, and trimer signals. With this in mind, Fig. \ref{fig3} shows the combined theoretical dimer and trimer spectrum (dot-dashed black curves) compared to experimental spectra (solid blue curves). 
The trimer signals are obtained by convoluting the histograms with Gaussian profiles having a width of 700 KHz for $34d_{3/2}$ and 3 MHz for $36d_{3/2}$. We normalize them to observed signals.
The addition of the trimer lines significantly improves agreement with the observed data. The broad peaks near $45$ and $80$ MHz detuning in the $34d_{3/2}$ spectrum illustrate the presence of non-additive trimer states where dimer lines are nearly absent. A similar trimer peak can be identified in the $36d_{3/2}$ spectrum, see red circle in Fig. \ref{fig3}.  Furthermore the higher resolution data for $34d_{3/2}$ reveals substructures in the trimer peaks (upper panel in Fig. \ref{fig3}) that are also predicted by our simulations and can be related to the energy spacing of the PESs of approximately 4 MHz visible in Fig. \ref{fig1} (c) which is only resolved at sufficiently small line widths. The additional structures seen in the experimental spectra below $-100$ MHz might be due to the presence of radial trimers associated to inner wells seen in Fig. \ref{fig2}. These states are influenced by the presence of the $p-$wave resonances and are not captured in our theory of three-body interactions.

We demonstrate that an effective three-body interaction can be realized by Rydberg excitation into anisotropic trimer ultralong-range molecules in an ultracold Cs gas. By including relativistic spin-dependent interactions in the Rydberg molecule Hamiltonian, we systematically analyze and identify observed spectroscopic features, as angular trimer Rydberg molecules whose energies are {\it not} multiples of dimer energies. Spectral features of radial dimers also are found and interpreted with accuracy. In the future,  electric or magnetic fields can be employed to resolve and identify individual molecular lines due to the effective three-body terms. An interesting question that might be answered with an increased accuracy is whether the van der Waals interaction between the ground state atoms can have measurable effects for molecular trimers.  

\begin{acknowledgments}
C.F. gratefully acknowledges a scholarship by the Studienstiftung des deutschen Volkes. P.S. acknowledges support from the Deutsche Forschungsgemeinschaft (DFG) within the Schwerpunktprogramm 1929 Giant Interactions in Rydberg Systems (GiRyd). J.Y., F.M. M.B. and J.P.S. acknowledge support from NSF Grant No. PHY-1607296. S.T.R. acknowledges support from NSF Grant No. PHY-1516421 and funding from the Research Corporation for Science Advancement. J.Y. lead the experimental effort. C. F. thanks T. Pfau, R. L\"ow and F. Meinert for fruitful discussions.
\end{acknowledgments}

  \input{bib.bbl}

\pagebreak
\onecolumngrid
\setcounter{equation}{0}
\setcounter{figure}{0}
\renewcommand{\theequation}{S.\arabic{equation}}
\renewcommand{\thefigure}{S\arabic{figure}} 
\section{Supplemental material}
\subsection{Electronic Hamiltonian}
We consider a triatomic Rydberg molecule that consists of a Rydberg atom (positively charged core with a Rydberg electron) and two ground state atoms. The relative distances of the electron and the $i$-th ground state atom to the core are given by $\vec{r}$ and $\vec{R}_i$, respectively. Our model for the electronic trimer Hamiltonian is
\begin{align}
\hat{H}= \hat{H}_0^\text{ryd} + \hat{H}_1^\text{HF} + \hat{H}_2^\text{HF} + \hat{V}_1 + \hat{V}_2 - \frac{\alpha}{2R^4_1} - \frac{\alpha}{2R^4_2} .
\label{eqn:Hamiltonian}
\end{align}
Here $\hat{V}_1$ and $\hat{V}_2$ are the Fermi pseudopotentials as described in the main part and $\hat{H}_0^\text{ryd}$ is the Hamiltonian of the unperturbed Rydberg atom. It takes into account spin-orbit coupling and has eigenstates $\ket{nljm_j}$ with energies
\begin{equation}
E_{nlj}=-\frac{1}{2 (n-\mu(n,l,j))^2} ,
\end{equation}
where $\mu(n,l,j)$ are quantum defects that can be determined experimentally \cite{goy_millimeter-wave_1982}. For $30 \leq n \leq 40$ one has $\mu(n,2,3/2)\approx 2.48$ and $\mu(n,2,5/2) \approx 2.47$.  
The radial Rydberg wave functions can be approximated by Whitaker functions that depend parametrically on the effective quantum number $n_\text{eff}=n+\mu(n,l,j)$ \cite{eiles_hamiltonian_2017}.

Furthermore, each ground state atom has an internal hyperfine structure described by
$\hat{H}_i^\text{HF}= A_\text{HF} \vec{I}_i \cdot \vec{s}_i$ with the nuclear spin $\vec{I}_i$ and the electron spin $\vec{s}_i$ which can be coupled to $\vec{F}_i= \vec{I}_i+ \vec{s}_i$. For $^{133}$Cs atoms one has $I_i=7/2$, $s_i=1/2$ and the hyperfine constant is $A_\text{HF}= 2.298 \text{ GHz}$.

Additionally, the model includes effects due to the polarization of the ground state atoms by the ionic core that lead to a $-\frac{\alpha}{R_i^4}$ interaction, where $\alpha=402.2$ a.u. is the polarizability of the $^{133}$Cs atom.

\subsection*{Pseudopotential with $p$-wave interactions}
The Hamiltonian given in (\ref{eqn:Hamiltonian}) can be readily specialized for the computation of diatomic potential energy curves by neglecting the terms $-\frac{\alpha}{2 R_2^4}$ and $\hat{V}_2$ (this corresponds to the limit $R_2 \to \infty$). In addition we consider also spin-orbit coupled $p$-wave interactions in $\hat{V_1}$ that take into account the coupling between the combined electronic spin $S$ (singlet or triplet) and the electronic angular momentum $L$ ($s$-wave or $p$-wave). An expression for this potential has been derived recently in \cite{eiles_hamiltonian_2017}.
It can be written in the form
\begin{equation}
\hat{V}= \sum_{\beta} \frac{(2L+1)^2}{2} a(S,L,J,k) \frac{\delta(X)}{X^{2(L+1)}} \ket{\beta} \bra{\beta}.
\label{eqn:pseudopotential_supp}
\end{equation}
Here $X=|\vec{r}-\vec{R}|$ is the relative distance between the Rydberg electron and the ground state atom at $\vec{R}$. The multiindex $\beta$ labels projectors onto different scattering channels that are characterized by collective electronic configurations of the ground state atom and the Rydberg electron $\ket{\beta}=\ket{L S J M_{J}}$. The quantum numbers $L$,$S$,$J$,$M_{J}$ specify the total orbital angular momentum $\vec{L}$ in the reference frame centered on the ground state atom ($L=0$ for $s$-wave and $L=1$ for $p$-wave scattering), the total spin $\vec{S}$ ($S=0$ for singlet and $S=1$ for triplet scattering), and the total angular momentum $\vec{J}=\vec{L} + \vec{S}$ ($J\in \{1,2,3\}$, $M_{J}\in\{-J_i,\dots, J_i\}$). The interaction strength in each channel is determined by the scattering lengths $a(S,L,J,k)$ for the ground state and a free electron with momentum $k$. We use here phase shift data identical to the one presented in \cite{markson_theory_2016}. The $s$-wave zero energy scattering lengths are $a(3,0,0,0)=-21.7 \text{ a.u.}$ and $a(1,0,0,0)=-1.33 \text{ a.u.}$. 

To evaluate the action of $\hat{V}$ onto the basis states one needs to know the matrix that mediates the transformation between the reference frame of the Rydberg atom and the reference frame of the ground state atom. This transformation matrix has been derived in \cite{eiles_hamiltonian_2017}. It is, however, only valid for a situation where the ground state atom lies on the $z$-axis. For our discussion of trimers we do therefore focus on a regime where only $s$-wave interaction is important and where $\vec{L} \cdot \vec{S}$ coupling is negligible.

\subsection{Electronic structure of the dimer potentials}
\begin{figure}[h]
\includegraphics[width= 0.4 \textwidth]{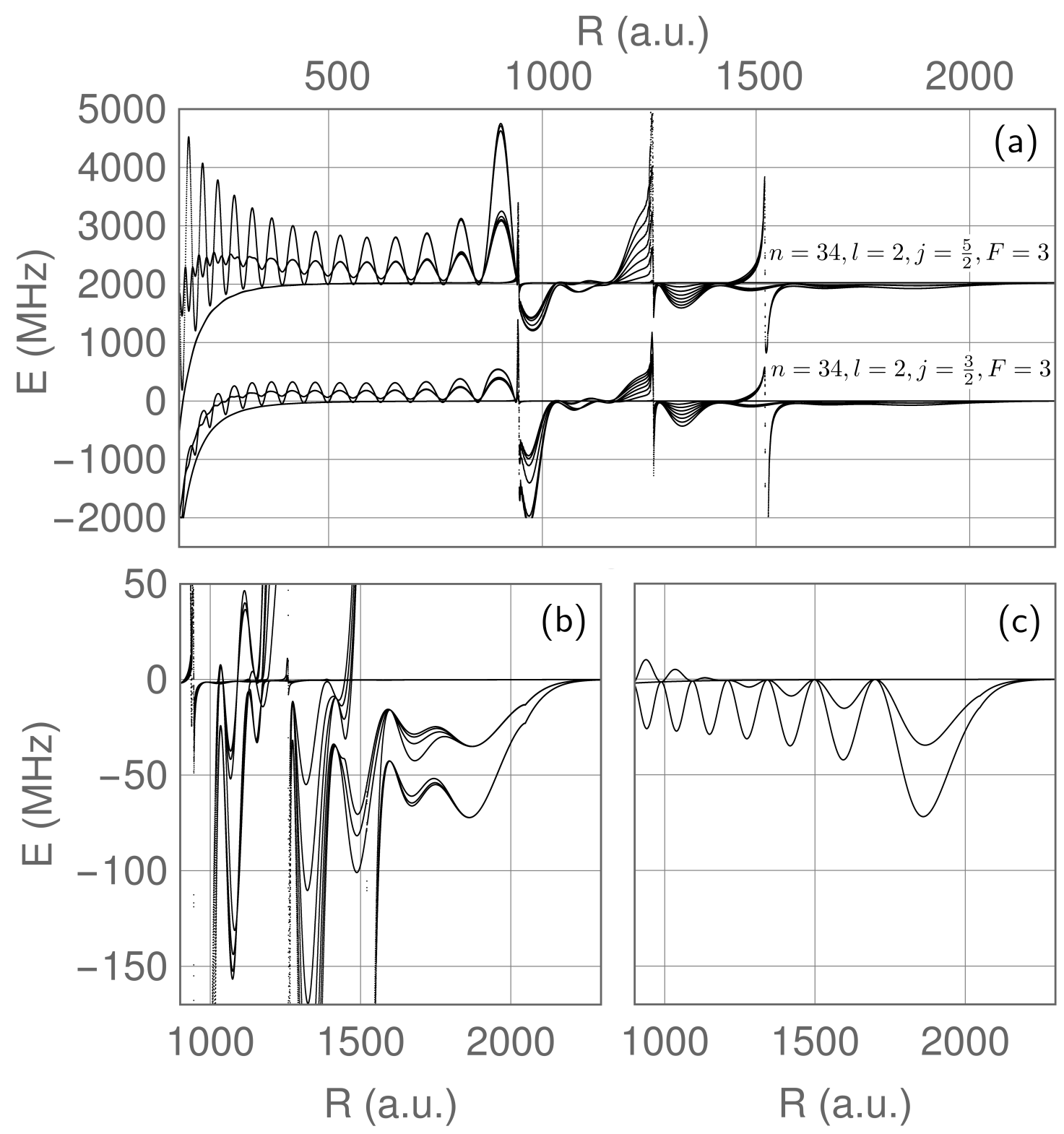}
\caption{Potential energy curves for diatomic Rydberg molecules with $p$-wave interaction (a), (b) and without $p$-wave interaction (c). The zero energy has been set to the energy of the $34d_{3/2}$, $F=3$ level.}
\label{fig:dimer_supp}
\end{figure}

To provide a better overview of the electronic structure we present in Fig. \ref{fig:dimer_supp} (a) the diatomic energy curves of a Cs(34$d_{3/2}$) Rydberg molecule close to the $F=3$ state on a scale where the neighboring $j=5/2$ level is visible. States with $F=4$ are not visible since, for $n=34$ the hyperfine splitting ($\sim$ 9 GHz) is large compared to the fine structure splitting ($\sim$2 GHz). The three poles around 900 a.u., 1300 a.u. and 1500 a.u. result from the resonances in the three triplet $p$-wave channels. For small distances ($R< 500 \text{ a.u.}$) the influence of the  $-\alpha/(2R^4)$ potential becomes visible.
Fig. \ref{fig:dimer_supp} (b) shows a magnified region below the $j=3/2$ which is identical to the 34$d$ PESs presented in Fig. 2 of the main article.
Neglecting the $p$-wave interaction leads to the simpler PESs displayed in Fig. \ref{fig:dimer_supp} (c). In this case, one can distinguish six (eight) nearly degenerate deep (shallow) potential curves, that can be characterized by the combined angular momentum $\vec{G}=\vec{s}_r+\vec{F}$ and its projection along the internuclear axis. The deepest curves ($G=5/2$) corresponds to nearly pure triplet character, while the shallow curves ($G=7/2$) have a mixed singlet/triplet character \cite{anderson_photoassociation_2014,sasmannshausen_experimental_2015,bottcher_observation_2016}. 
In Fig. \ref{fig:dimer_supp} (b) the 8-fold (6-fold) degeneracy of the shallow (deep) potential curves becomes lifted for inner wells due to the $p$-wave interaction and each curve splits into different curves of constant $m_j+m_{F_1}$ (which is the conserved projection of the total orbital momentum along the internuclear axis) \cite{eiles_hamiltonian_2017}.

\subsection{Electronic structure of the trimer angular potentials}

\begin{figure*}[t]
\includegraphics[width= 0.9 \textwidth]{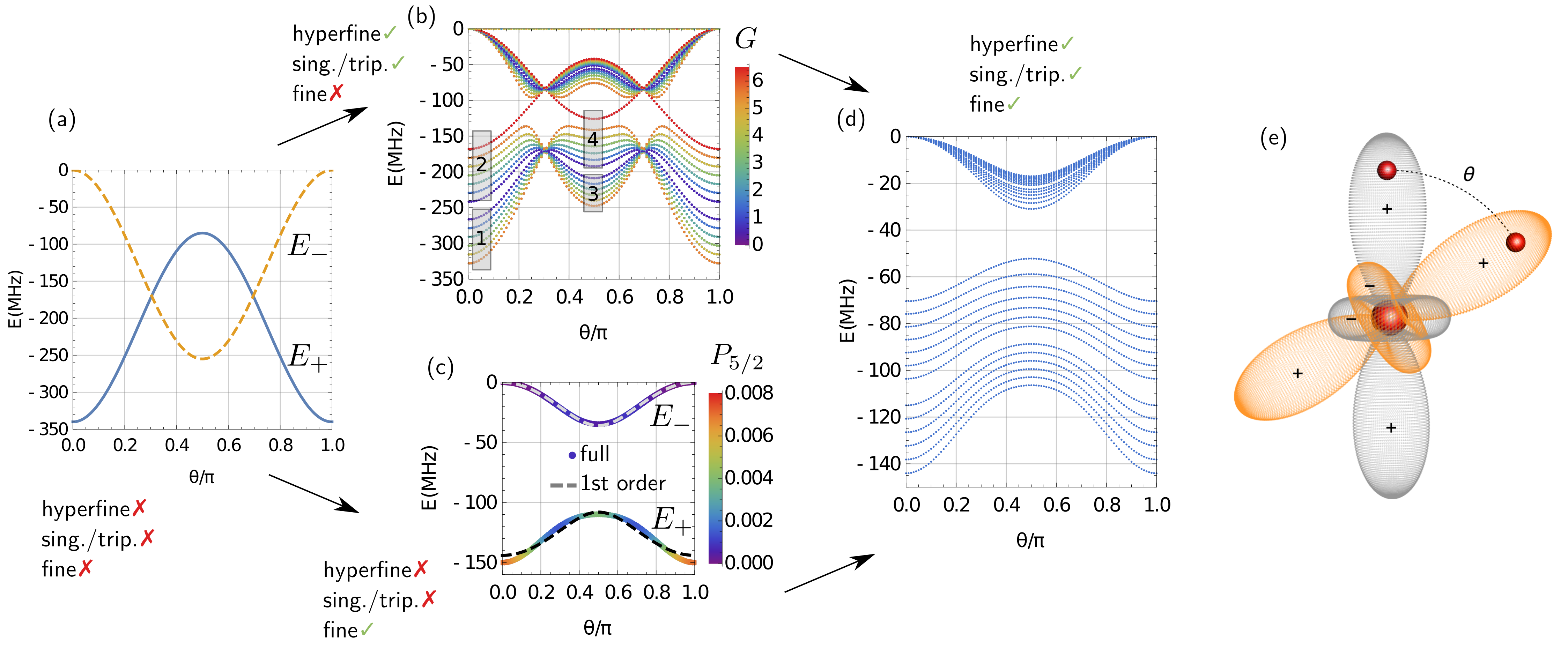}
\caption{Cut of the $34d$ trimer PESs for fixed $R_1=R_2=1868$ and variable $\theta$. The different spin interactions (hyperfine interaction, singlet/triplet splitting, Rydberg finestructure) are added successively from (a) to (d). The
zero energy has been set to the energy of the $34d$, $j=3/2$, $F_1=F_2=3$ state. In (b) the color encodes the value of the good quantum number $G$ (see text) and the gray boxes divide the PESs into the four groups $E_{+\text{deep}}$ (1), $E_{+\text{shallow}}$ (2), $E_{-\text{deep}}$ (3) and $E_{-\text{shallow}}$ (4). The color in (c) encodes the admixture of $j=5/2$ character $P_{5/2}$. The orbitals in (e) illustrate the angular distribution (including sign changes) of the two diatomic orbitals $\psi_\text{dim}(\vec{r};\vec{R}_{1/2})$ whose symmetric and antisymmetric superpositions $\psi_\pm(\vec{r};\vec{R}_1;\vec{R}_2)$ form the wave function of the triatomic molecule (without any spin effects). The angular configuration of the Rydberg core and the two ground state atoms is represented by the central sphere and the outer spheres, respectively.}
\label{fig:structure_supp}
\end{figure*}

While the two-state model in Fig. \ref{fig:structure_supp} (a) explains the main binding mechanism of Cs($nd$) Rydberg trimers, it doesn't allow for a quantitative agreement with experimental observations. It neglects the hyperfine structure of the ground state Cs atoms, the fine structure of the Rydberg atom, nor does it distinguish between singlet and triplet scattering. The spin-insensitive scattering is modeled with the energy dependent triplet $s$-wave scattering length of the complete system. 
Including all of these additional effects gives rise to the angular PESs in Fig. \ref{fig:structure_supp} (d). To unravel how their complex structure depends on the interplay of the various spin-effects we will switch on/off selected interactions successively and, thus, trace the transition from the simple two-state model from Fig. \ref{fig:structure_supp} (a) to  Fig. \ref{fig:structure_supp} (d).     

Firstly, as an academic exercise, we turn on the Rydberg spin-orbit coupling, but neglect the hyperfine interaction. The PESs for $j=3/2$ are shown in Fig. \ref{fig:structure_supp} (c). The crossing degeneracies of the $E_\pm$ curves in Fig. \ref{fig:structure_supp} (a) (see also main text) are lifted, but the general form of the PESs remains similar. The shape of these curves can be approximated via first order perturbation theory in the limit that the fine structure splitting between the $j=3/2$ and the $j=5/2$ state is large compared to the depth of the PESs to
\begin{equation}
E_\pm (R,\theta)\approx  E_\text{dim}(R) \left(1 \pm \frac{\sqrt{10+6 \cos 2 \theta}}{4}\right).
\label{eqn:E_justfine}
\end{equation}
Deviations between these approximate curves (\ref{eqn:E_justfine}) and the numerical results which are visible in Fig. \ref{fig:structure_supp} (c) are due to higher order effects. In particular the $s$-wave interaction admixes a certain amount of $j=5/2$ character to the predominately $j=3/2$ states which is indicated by the color encoding.

Secondly, the hyperfine interaction in the ground state Cs atoms is turned on and we allow for scattering in singlet and triplet channels (the fine structure is, however neglected), see Fig. \ref{fig:structure_supp} (b). 
The hyperfine interaction mixes singlet and triplet states and leads to a splitting of the ''spin-free`` PESs. Around the $\theta=0$ and $\theta=\pi$ minima (the $\theta= \pi/2$ minima) one finds 13 non-degenerate curves that originate from the $E_+$ ($E_-$) curve. These curves can be characterized by their approximately conserved total angular momentum $\vec{G}^2=(\vec{F}_1+\vec{F}_2+ \vec{S}_0)^2=G(G+1)$ (see color encoding in \ref{fig:structure_supp} (b)), where $G$ takes half-integer values between $13/2$ and $1/2$. Consequently each of these curves belongs to a number of ($2G$) degenerate states.
The 13 curves can be further subdivided into 6 deeper curves $E_{\pm \text{deep}}$ and 7 shallower curves $E_{\pm \text{shallow}}$ as indicated by the rectangular regions in Fig. \ref{fig:structure_supp} (b). This splitting is similar to the dimer case where one distinguishes deep triplet dominated curves and  shallower curves of mixed singlet/triplet character, see Fig \ref{fig:dimer_supp} (b).

Finally, the PESs in Fig. \ref{fig:structure_supp} (d) combine all the discussed aspects: The equilibrium angles result from the structure of the $l=2$ orbitals, the splitting is a consequence of the hyperfine and singlet/triplet couplings, while the avoided crossing can be linked to the Rydberg spin-orbit coupling.

\subsection{Supplemental information on the experiment}
The lasers used for the crossed dipole trap have beam waists of 98 $\mu$m and the aspect ratio of the crossing is approximately 2:1. The trap frequencies are $2\pi \cdot 3.58 \text{ KHz}$ along the short axis and $2\pi \cdot 1 \text{ KHz}$ along the long axis. This corresponds to a trap depth of 5 mK.

For the excitation we use a 852 nm laser with a spot size of 1 mm diameter and a power of 5 mW, while the 508 nm laser has a spot size of 25 micron diameter and a power of 130 mW (140 mW) for the $34d_{3/2}$ ($36d_{3/2}$) measurement. The overall laser stability is 700 KHz (3 MHz) for the $34d_{3/2}$ ($36d_{3/2}$) spectra. Furthermore a better signal to noise ratio was achieved for the $34d_{3/2}$ measurements by measuring both, the Cs$^+$ and the Cs$_2^+$ signal.

\begin{figure}[h]
 \includegraphics[width= 0.8 \textwidth]{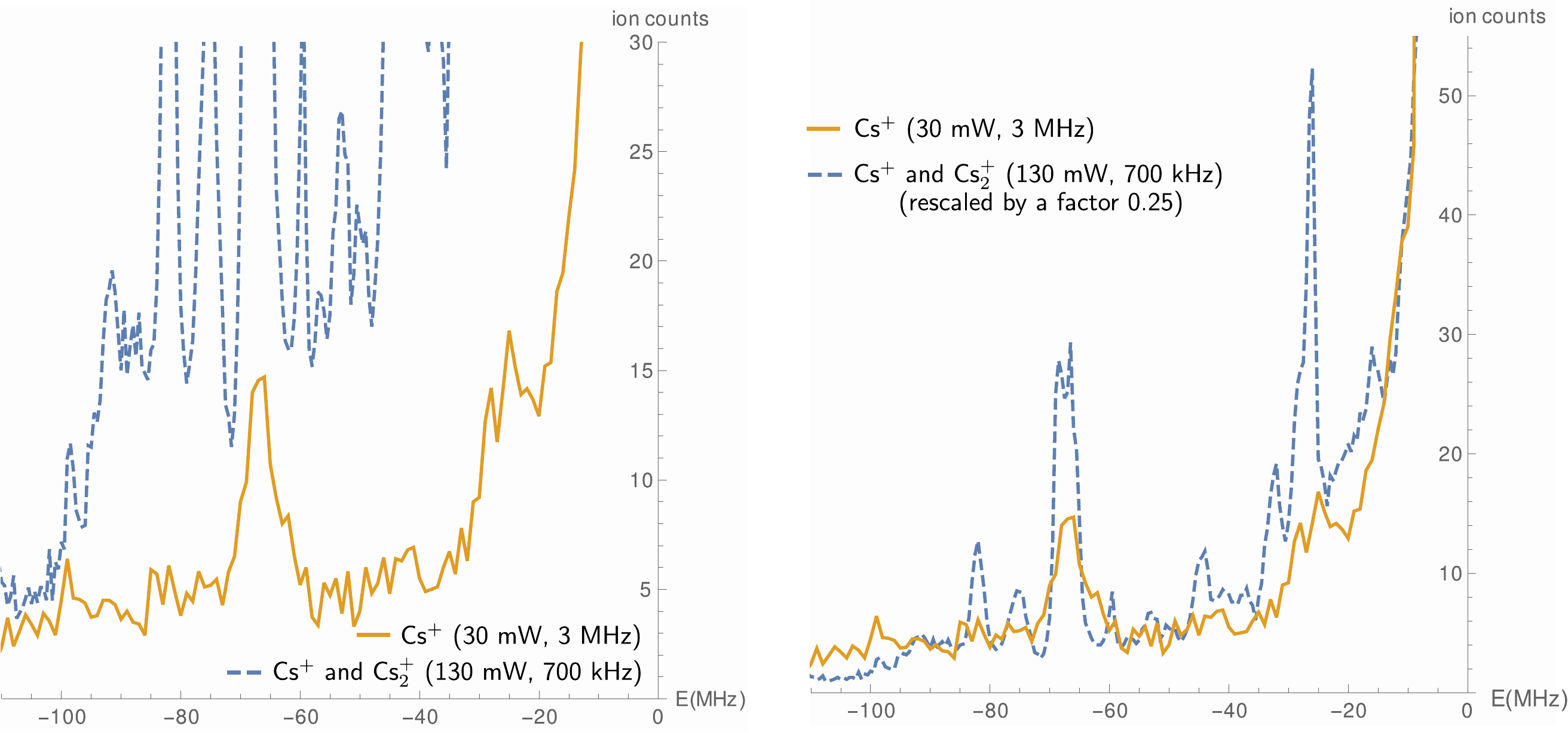}
\caption{Experimental ion signal as a function of the laser detuning from the $34d_{3/2}$ Rydberg line for different laser power and different line widths. Left: The higher resolution signal shows the combined Cs$^+$ and Cs$_2^+$ signal (dashed line), whereas the lower resolution signal contains only the Cs$^+$ signal. Right: Here the higher resolution spectrum (dashed line) is rescaled by a factor of 0.25.}
\label{fig:linewidth_comparison_supp}
\end{figure}

To illustrate the impact of the line width and the detection method we present in Fig.~\ref{fig:linewidth_comparison_supp} a comparison between two experimental spectra close to the $34d_{3/2}$ Rydberg line. One spectrum is the combined Cs$^+$ and Cs$_2^+$ signal which is also presented in the main part of the manuscript. The other spectrum shows only the Cs$^+$ signal and was obtained with a larger line width (3 MHz instead of 700 KHz).
It becomes evident that every peak in the high resolution spectrum has a counterpart in the low resolution spectrum. However, the visibility of these peaks suffers from the lower resolution and the lower signal to noise ratio. 
While both spectra capture the main features of the trimer peaks (asymmetric line shape and shoulders), only the high-resolution data resolves clearly the individual PES visible in Fig. \ref{fig:structure_supp} (d). 

\subsection{Additional Cs($nd_{3/2}$) molecular spectra and simulations}
\begin{figure*}[h]
 \includegraphics[width= 0.45 \textwidth]{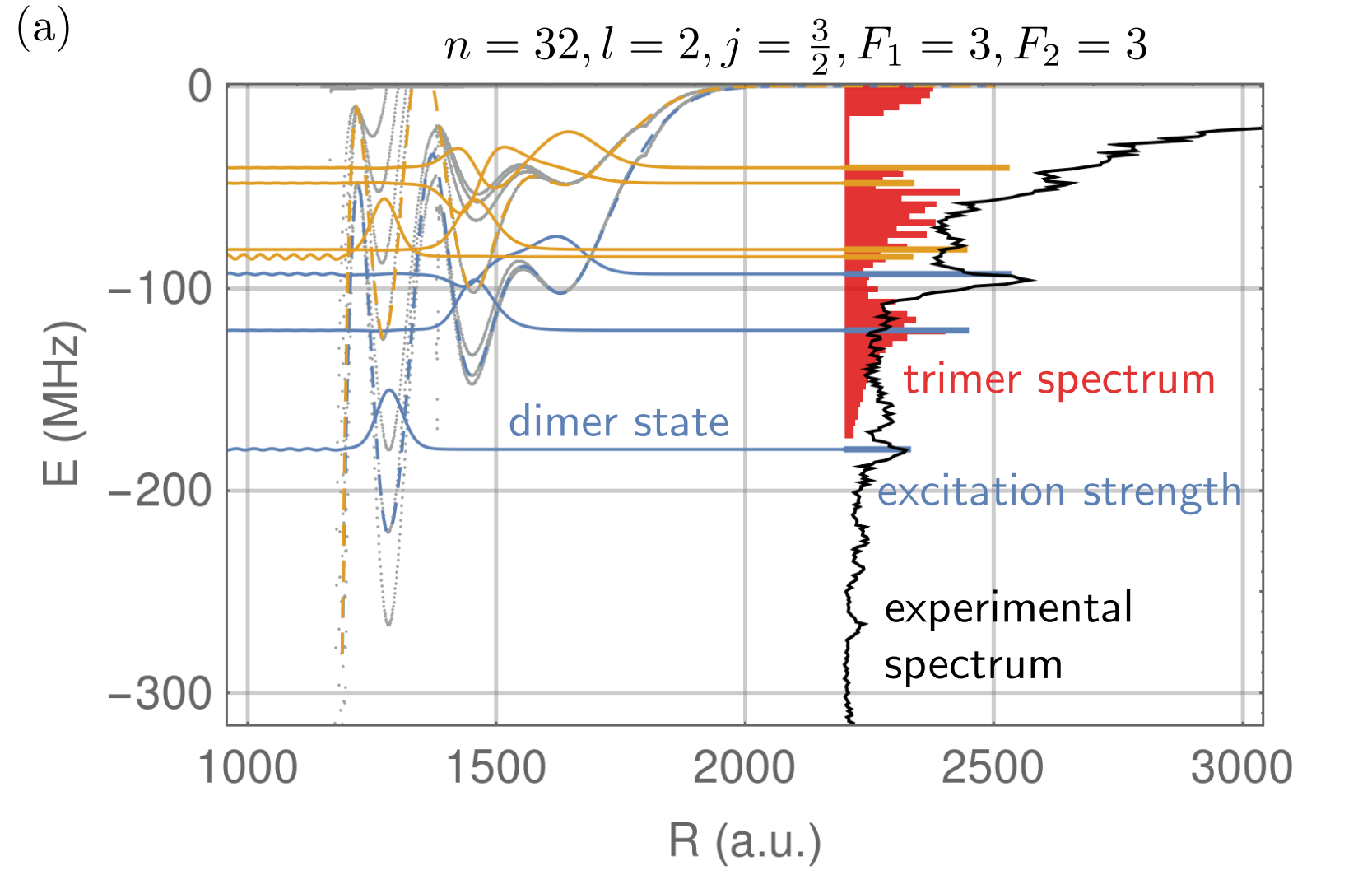}
\includegraphics[width= 0.45 \textwidth]{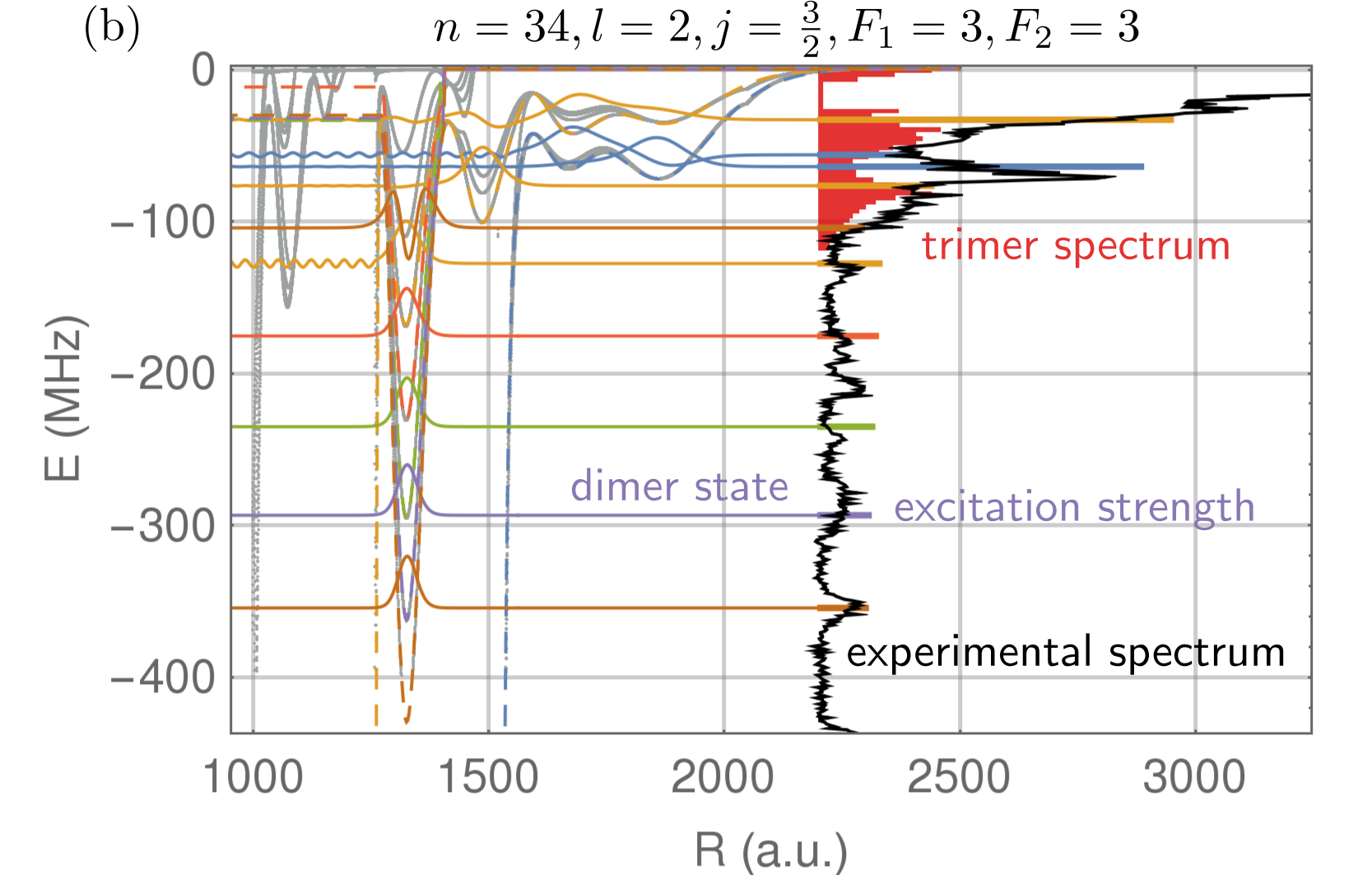} \\
\includegraphics[width= 0.45 \textwidth]{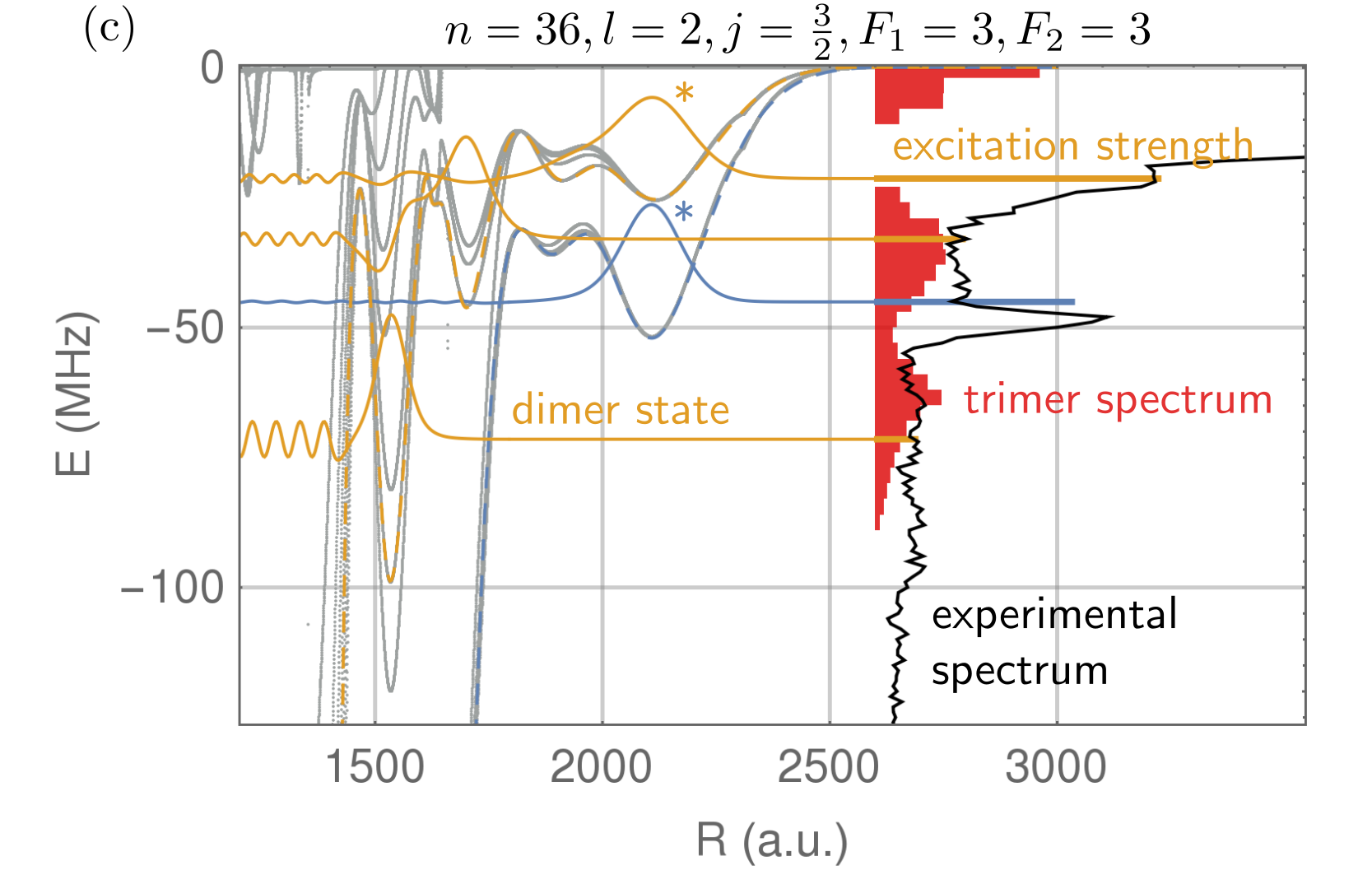} 
\includegraphics[width= 0.45 \textwidth]{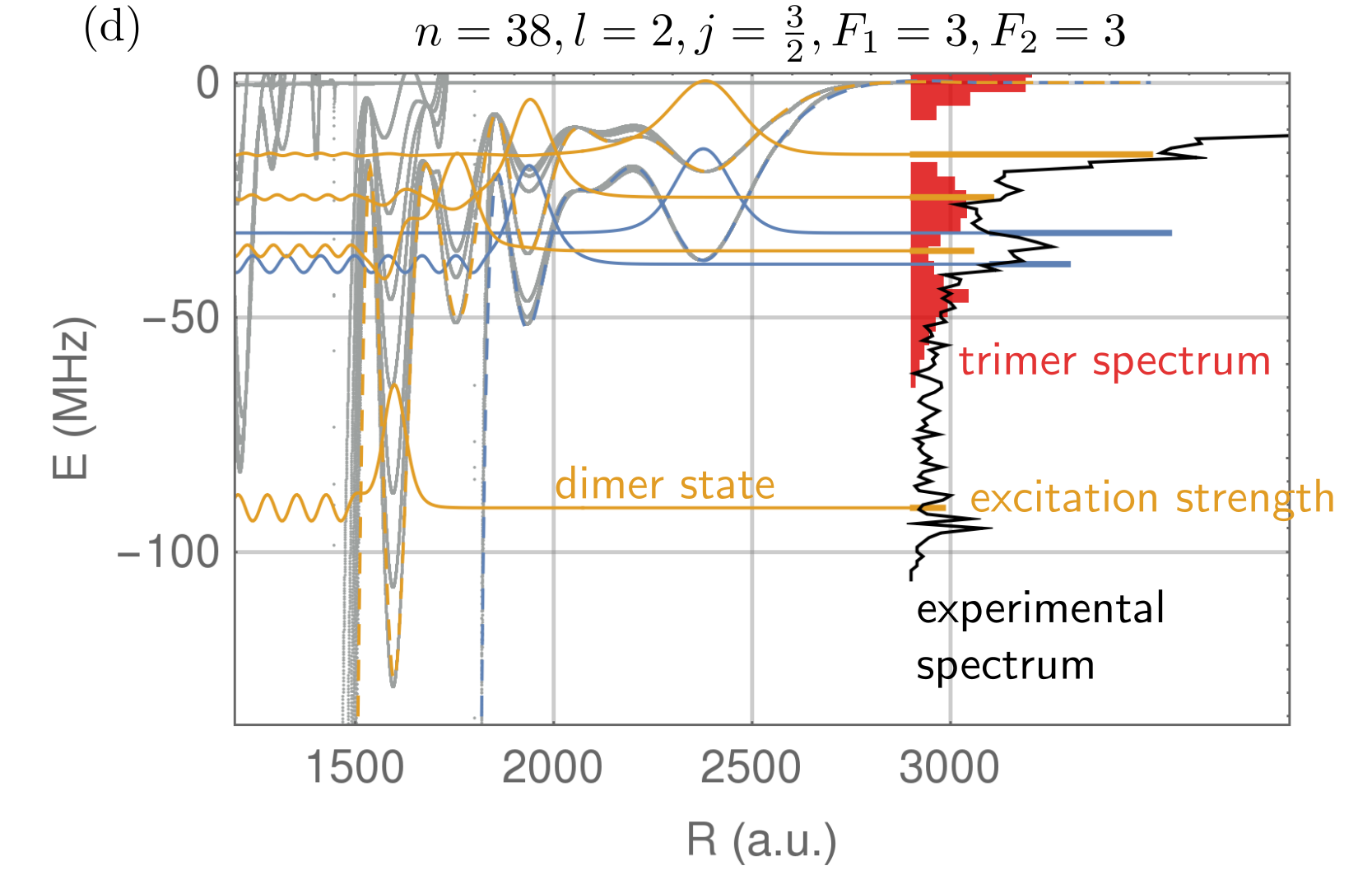}
\caption{Comparison between the experimental spectrum (black line) and computational results for molecules close to the states $32d_{3/2}$, $F=3$ (a), $34d_{3/2}$, $F=3$ (b), $36d_{3/2}$, $F=3$ (c) and $38d_{3/2}$, $F=3$ (d). Vibrational wave functions (colored lines) are presented for a few diatomic curves (dashed lines in the same color) that have been selected from the complete potential energy curves (gray background). The offset of the wave function corresponds to its vibrational energy. The colored bars on top of of the experimental spectrum represent the theoretical line strengths. The computed trimer spectrum is superimposed in red and includes a blue-shift which encodes the zero-point energy of stretching motions. The laser power is 140 mW in (a) and (b) and 30 mW in (c) and (d).}
\label{fig:spec_supp}
\end{figure*}

In Fig. \ref{fig:spec_supp} we show additional experimental spectra and theoretical simulations for the states $n=32$, $n=34$, $n=36$ and $n=38$ that have not been presented in the main part of the manuscript. The experimental spectrum for $n=34$ in \ref{fig:spec_supp}  (b) was obtained at a higher laser power (140 mW instead of 30 mW) and contains signals of deeply bound states at energies below -200 MHz. The comparison to our numerical data suggests that some of these signals stem from dimers that are bound in an inner well whose depth results from an avoided crossing with other curves due to the resonant $p$-wave interaction. The experimental spectrum for $n=36$ in Fig.~\ref{fig:spec_supp} (c) was obtained at a lower laser power than the corresponding spectrum in the main part (30 mW instead of 140 mW) and does not contain additional structures below 100 MHz. 
Similar to the histograms presented in the main text, we include a blue-shift which accounts for the zero-point energy of the molecular stretching motions.
This shift is 30 MHz, 25 MHz, 15 MHz and 10 MHz for $n=32$, $n=34$, $n=36$ and $n=40$, respectively. The shift corresponds approximately to two times the zero point energy of the dimer states in the most outer wells and, accordingly, decreases with increasing $n$. For $n=36$ the corresponding dimer states are marked with a star in Fig. ~\ref{fig:spec_supp}~(c).

\end{document}

%% file: bib.bbl
%